\documentclass[numbers,sort&compress]{elsarticle}

\usepackage{amsmath}
\usepackage{epsf}
\usepackage{graphicx}
\usepackage{dcolumn}
\usepackage{bm}

\def\br{{\bf r}}
\def\bJ{{\bf J}}
\def\bF{{\bf F}}
\def\cL{{\cal L}}

\newcommand{\beq}{\begin{equation}}
\newcommand{\eeq}{\end{equation}}
\newcommand{\be}{\begin{equation}}
\newcommand{\ee}{\end{equation}}
\newcommand{\beqa}{\begin{eqnarray}}
\newcommand{\eeqa}{\end{eqnarray}}
\newcommand{\bea}{\begin{align}}
\newcommand{\eea}{\end{align}}
\usepackage{amssymb}
\usepackage{array}
\usepackage{amsmath}
\usepackage{graphicx}\usepackage{graphics}
\usepackage{dcolumn}
\usepackage{bm}\usepackage{varioref}

\usepackage{amsmath}
\usepackage{epsf}
\usepackage{graphicx}

\usepackage{subfigure}
\usepackage{dcolumn}
\usepackage{bm}

\usepackage{amssymb}
\usepackage{array}
\usepackage{varioref}
\usepackage{wrapfig}
\usepackage{etoolbox}
\usepackage{color}

\def\prb{Phys. Rev. B }
\def\prl{Phys. Rev. Lett. }

\def\ie{{\it i.e.}}
\def\eg{{\it e.g.}}

\def\bnabla{\bm{\nabla}}

\def\br{{\bf r}}

\def\bJ{{\bf J}}
\def\bF{{\bf F}}

\def\ea{{e_\alpha}}

\def\be{\begin{equation}}
\def\ee{\end{equation}}
\def\ba{\begin{eqnarray}}
\def\ea{\end{eqnarray}}

\providecommand{\br}{\bv{r}}

\makeatletter
\newcommand*{\balancecolsandclearpage}{%
  \close@column@grid
  \clearpage
  \twocolumngrid
}
\makeatother

\begin{document}

\title{New exact results for the two-phase model with several conserved currents}

\author[a]{Oded Agam}
\address[a]{The Racah Institute of Physics, The Hebrew University of Jerusalem, 91904, Israel}
\author[b,c]{Igor L. Aleiner}
\address[b]{Physics Department, Columbia University, New York, NY 10027, USA}
\address[c]{Google, Venice, CA 90291, USA}
\author[a]{Dror Orgad}
\date{\today}
\begin{abstract}

We consider the macroscopic transport properties of two-dimensional random binary mixtures
with identical spatial distributions of the two phases. Previous studies have obtained
exact analytical results for the electrical conductivity of a single layer with and without
a magnetic field, as well as for the thermoelectric response of a magnetic field-free double-layer.
Here, we generalize these exact solutions to the magneto-thermoelectric response of a single layer
and to the thermoelectric response of a double-layer. The magneto-thermoelectric transport coefficients
of the double-layer are calculated perturbatively for weak magnetic field.

\end{abstract}


\maketitle

\section{Introduction}

Random mixtures of several phases are common in nature. They appear, for example, near first order phase transitions,
in percolating systems and when competing short and long-range interactions are present \cite{Seul}. Here, we are concerned
with the situation where the inhomogeneity occurs on sufficiently large scales, such that each phase can be
characterized by its own bulk kinetic properties, \eg, its electrical and thermal conductivities. Typically,
the calculation of the response of the entire system is intractable due to the complicated distribution of
currents in the sample. However, an exact analytical solution to this problem exists for two-dimensional systems
comprised of two phases with statistically identical distributions, as occurs at the percolation transition point.
The solution relies on the existence of a self-duality transformation, which interchanges the roles of the currents
and the driving fields. The duality can be realized provided that the fields are potential gradients and the currents
are conserved.

The aforementioned approach is originally due to Dykhne, who applied it to the case of a single (electrical) current
\cite{Dykhne71}. Subsequently, it was generalized to include the effects of a magnetic field
\cite{Dykhnemag,Shklovskii,Milton,DR1994,Balagurov-galvano95},
and to the case of two conserved currents (electrical and heat) but without a magnetic field \cite{Balagurov-reciprocity}.

In this paper we present several new exact solutions to the problem. First, we study a time-reversal symmetric system with
three conserved currents. Such a setting applies when measuring drag in a bilayer of electrically isolated films that are
strongly coupled thermally. Second, we provide a solution to the case of two conserved currents in the presence of a magnetic
field, $H$. Finally, we return to the bilayer problem and solve it perturbatively in $H$.

\section{Statement of the Problem}

Consider a two-dimensional system composed of two isotropic phases whose random
spatial distributions are statistically equivalent. Assume that there is a set of $N$
conserved current densities in the problem ${\cal J}=(\bJ_1,\cdots,\bJ_N)^T$, with
\begin{equation}
\label{eq:contJ}
\bnabla\cdot\bJ_\alpha=0,
\end{equation}
that can be induced by a set of $N$ external forces, ${\cal F}=(\bF_1,\cdots,\bF_N)^T$,
satisfying
\begin{equation}
\label{eq:Fcond}
\bnabla\times\bF_\alpha=0.
\end{equation}
Within linear response the currents and forces are
related by
\begin{equation}
\label{linearres}
{\cal J}(\br)=\hat L(\br){\cal F}(\br),
\end{equation}
where $\hat L(\br)$ is a matrix containing the kinetic coefficients, which takes
one of two values $\hat L_1$ or $\hat L_2$ corresponding to the phase present at point $\br$.
We choose the forces such that the local entropy production rate is given by ${\cal J}(\br)\cdot{\cal F}(\br)/T(\br)$.
Consequently, Onsager's relations and the isotropy of the constituent phases assure that in the absence
of a magnetic field $\hat L$ is symmetric.

Our goal is to calculate the macroscopic response of the system, as given by $\hat L_{\rm eff}$
relating the spatial averages of ${\cal J}$ and ${\cal F}$
\begin{equation}
\label{Leffdef}
\langle{\cal J}\rangle = \hat L_{\rm eff}\langle {\cal F}\rangle.
\end{equation}
To this end, we follow Dykhne \cite{Dykhne71} and introduce an auxiliary transport problem
defined in terms of
\begin{eqnarray}
\label{dualtrans}
&&{\bf J}'_\alpha= U_{\alpha\beta}\, \hat{\bf n} \times {\bf F}_\beta,~~~~~{\bf F}'_\alpha =U^{-1}_{\alpha\beta}\,
\hat{\bf n} \times {\bf J}_\beta,
\end{eqnarray}
where $\hat U$ is a symmetric matrix whose components are of the same physical dimensions as the components of $\hat L$,
$\hat{\bf n}$ is a unit vector perpendicular to the plane and where repeated indices are summed over.
One can easily check that the new currents and forces satisfy the requirements $\bnabla\cdot\bJ'_\alpha=0$ and
$\bnabla\times\bF'_\alpha=0$. Furthermore, owing to the linear nature of the transformation (\ref{dualtrans})
they are related by ${\cal J}'(\br)=\hat L'(\br){\cal F}'(\br)$. From Eqs. (\ref{linearres})-(\ref{dualtrans})
it follows that $\hat L'(\br)=\hat U \hat L^{-1}(\br) \hat U$ and
\be
\label{Leffrelation}
\hat L'_{\rm eff}=\hat U \hat L_{\rm eff}^{-1} \hat U.
\ee
A particularly useful choice for $\hat U$ is the one that interchanges the two components, {\it i.e.},
\be
\hat L_2 = \hat U \hat L_1^{-1} \hat U. \label{rotation}
\ee
For such a duality transformation the auxiliary problem corresponds to a system that is obtained from the original
one by replacing one phase by the other. Consequently, their statistical equivalence implies that
$\hat L'_{\rm eff}=\hat L_{\rm eff}$, which together with Eq. (\ref{Leffrelation}) leads to $\hat L_{\rm eff}=\hat U$,
and thus
\be
\label{selfdualrel}
\hat L_2 = \hat L_{\rm eff}\hat L_1^{-1} \hat L_{\rm eff}.
\ee
This algebraic equation determines the macroscopic response of the system.

\subsection{Examples}
In the case of a single conserved (electrical) current the matrices $\hat L_{1,2}$ become the conductivities
of the two phases, $\sigma_{1,2}$, and Eq. (\ref{selfdualrel}) reduces to $\sigma_2=\sigma_{\rm eff}^2\sigma_1^{-1}$
with the solution \cite{Dykhne71}
\be
\sigma_{\rm eff} = \sqrt{\sigma_1 \sigma_2}.
\ee

For two or more conserved currents the non-commutativity of $\hat L_1$ and $\hat L_2$ plays an essential role.
As an example, consider the thermoelectric response of a two-dimensional film described by
\be
\label{thermolinear}
\left(\begin{array}{c} {\bf j}(\br) \\{\bf q}(\br)  \end{array}\right)
= \hat L({\bf r}) \left(\begin{array}{c} {\bf E}(\br) \\-\frac{\boldsymbol{\nabla} T(\br)}{T(\br)} \end{array} \right),
\ee
where ${\bf j}$ and ${\bf q}$ are the electrical and heat current densities, respectively, while ${\bf E}$ and $T$
are the electric field and the temperature. Note that $\bf q$ is conserved within linear response that neglects
second order Joule heating effects. The solution to Eq. (\ref{selfdualrel}) in this case is \cite{Balagurov-reciprocity,Snarskii-arxiv}
\be
\label{LefflayerB0}
\hat L_{\rm eff}=c\left( \frac{\hat L_1}{\sqrt{d_1}}+\frac{\hat L_2}{\sqrt{d_2}}\right),
\ee
where $d_j=\det \hat L_j$ and the constant $c$ is
\be
\label{cdef}
c= (d_1 d_2)^{1/4}\left[\det\left( \frac{\hat L_1}{\sqrt{d_1}}+\frac{\hat L_2}{\sqrt{d_2}}\right)\right]^{-1/2}.
\ee
For completeness, we present the derivation of this result in the Appendix.

\section{Thermoelectric response of a two-phase double-layer}

Next, we consider the case with three conserved currents, motivated by drag experiments in double-layer
graphene \cite{exp1,exp2,exp3,exp4}. In these experiments the two graphene layers are electrically isolated
from each other, but they are in sufficient proximity such that they may be considered as a single layer
from a thermal point of view. This thermalization is primarily due to intra and inter-layer
electron-electron inelastic scattering, while the electron-phonon coupling is much weaker and does not
lead to a significant violation of the presumed conservation of heat in the system.
Under such conditions the linear response is described by
\be
\left(\begin{array}{c} {\bf j}_u(\br)\\  {\bf j}_d(\br) \\{\bf q}(\br)  \end{array}\right)= \hat L(\br)
\left(\begin{array}{c} {\bf E}_u(\br)\\  {\bf E}_d(\br)\\-\frac{\boldsymbol{\nabla}T(\br)}{T(\br)} \end{array} \right),
\ee
where ${\bf j}_{u,d}$ are the electrical current densities in the upper and lower layers, respectively,
and ${\bf q}$ is the total heat current density through the system.  Similarly, ${\bf E}_{u,d}$ are the electric fields
in the two layers, and $T$ is the temperature field, which is identical in both.
Furthermore, we will study the case where a domain of a particular phase in the upper layer appears above a similar
domain in the lower layer. As a result, $\hat L(\br)$ can take two values
\be
\label{eq:doublelayer}
\hat L_j= \left(\begin{array}{ccc}
\sigma_j & \eta_j & \alpha_j  \\
\eta_j & \sigma_j &  \alpha_j \\
\alpha_j &  \alpha_j  & 2\kappa_j  \end{array} \right), ~~~~~~ j=1,2
\ee
where $\sigma$ is the electrical conductivity, $\alpha/T\sigma$ is the thermopower and $\kappa/T$ is the
thermal conductivity (provided it is much larger than $\alpha^2/T\sigma$, as in metals \cite{Abrikosovbook}).
Finally, $\eta$ is the drag conductivity due to interaction-induced momentum transfer between the layers.

Beyond momentum transfer there is another drag mechanism which originates
from inter-layer energy transfer \cite{lev1,lev2,lev3}. Interfaces between phases act as thermocouples.
A current driven through one layer generates local temperature gradients by the Peltier effect. The strong
thermal coupling in the system causes these temperature gradients to propagate to the second layer,
which is under open circuit conditions, and generate thermopower there by the Seebeck effect.

Equation (\ref{Leffrelation}) can be exactly solved for the model defined by Eq. (\ref{eq:doublelayer}),
as described in the Appendix. Here, we focus on
two relevant special cases and begin by considering a system for which $\eta_1=\eta_2=0$, where any resulting
drag is due to inter-layer energy transfer alone.
Solving Eq. (\ref{selfdualrel}) we obtain the effective conductivity
\be
\sigma_{\rm eff}= \frac{1}{2} \sqrt{\sigma_1 \sigma_2} \left( 1+ \frac{ \nu^{1/2}\cos\theta_1+ \nu^{-1/2}\cos\theta_2}
{\sqrt{\nu+ \nu^{-1}+2\cos(\theta_1 + \theta_2)}}\right),
\ee
with
\be
\sin \theta_j=\frac{ \alpha_j}{\sqrt{\kappa_j \sigma_j}},~~~~~~ \nu= \frac{\sigma_1 \kappa_2}{\sigma_2 \kappa_1},
\ee
where the angles $\theta_j$ are defined in the range $[-\pi/2,\pi/2]$. The other effective transport coefficients
are given in terms of $\sigma_{\rm eff}$ according to
\begin{eqnarray}
&&\eta_{\rm eff} = \sigma_{\rm eff} -\sqrt{\sigma_1 \sigma_2}, \\
&&\alpha_{\rm eff}= \frac{\alpha_1 \sqrt{d_2}+\alpha_2 \sqrt{d_1}}{\sigma_1\sqrt{d_2}+ \sigma_2\sqrt{d_1}}
\left(2 \sigma_{\rm eff}-\sqrt{\sigma_1 \sigma_2} \right), \\
&&\kappa_{\rm eff} = \frac{\kappa_1 \sqrt{d_2}+\kappa_2 \sqrt{d_1}}{\sigma_1\sqrt{d_2}+ \sigma_2\sqrt{d_1}}
\left(2 \sigma_{\rm eff}-\sqrt{\sigma_1 \sigma_2} \right),
\end{eqnarray}
where, as before, $d_{1,2}=\det \hat L_{1,2}$.

Next, we treat the case with inter-layer momentum transfer but where the two phases differ only by the sign
of their charge carriers, while all other characteristics such as mobility and mass are identical, \ie,
\be
\label{Lphmodelbi}
\hat L_{1,2}= \left(\begin{array}{ccc}
\sigma & \eta & \pm\alpha  \\
\eta & \sigma & \pm\alpha \\
\pm\alpha & \pm\alpha & 2\kappa  \end{array} \right).
\ee
For this model we find
\begin{eqnarray}
\label{Leff0i}
&&\sigma_{\rm eff}= \sigma \cos^2 \left( \frac{\theta}{2} \right) - \eta \sin^2 \left( \frac{\theta}{2} \right) , \\
&&\eta_{\rm eff}=-\sigma \sin^2 \left( \frac{\theta}{2} \right) + \eta \cos^2 \left( \frac{\theta}{2} \right), \\
&&\alpha_{\rm eff}=0, \\
\label{Leff0f}
&&\kappa_{\rm eff}=\kappa \cos\theta,
\end{eqnarray}
where
\be
\sin\theta=\frac{\alpha}{\sqrt{\kappa (\sigma+\eta)}}.
\ee

\section{A system in a magnetic field}

Our next goal is to extend the discussion to include a magnetic field, $H$, along the $\hat z$ direction.
In the presence of the field the components of $\hat L(\br)$ and $\hat L_{\rm eff}$ in
Eqs. (\ref{linearres},\ref{Leffdef}) become $2\times 2$ tensors, whose off-diagonal terms are antisymmetric owing to the
assumption of isotropic phases. For example, the electrical conductivity tensor of the $j=1,2$ phases is
\be
\label{eq:sigform}
\hat\sigma_j =
\left(\begin{array}{cc}
\sigma_j & \sigma_{Hj}\\
-\sigma_{Hj} & \sigma_j\\ \end{array}\right),
\ee
where $\sigma_H$ is the Hall conductivity.

Equations (\ref{linearres}) and (\ref{Leffdef}) may be condensed by representing two-dimensional vectors
as complex numbers, {\it e.g.}, ${\bf E}=E_x+iE_y$, \cite{Balagurov-galvano95} leading to current densities
and forces of the form
\ba
\label{eq:vecsdef}
&&{\cal J}=\left( J_{1x}+iJ_{1y},\cdots,J_{Nx}+iJ_{Ny} \right)^T, \\
&&{\cal F}=\left( F_{1x}+iF_{1y},\cdots,F_{Nx}+iF_{Ny} \right)^T.
\ea
In this representation Eq. (\ref{linearres}) becomes
\be
\label{eq:complexnot}
{\cal J}(\br)=\hat{\cL}(\br){\cal F}(\br),
\ee
where the complex response matrix $\hat\cL(\br)$ takes one of two values in the corresponding phases $j=1,2$,
\be
\label{eq:Ldef}
\hat \cL_j = \hat L_j-i\hat L_{Hj},
\ee
expressed in terms of $N\times N$ symmetric matrices $\hat L_j$ and $\hat L_{Hj}$
holding the longitudinal and Hall parts, respectively, of the kinetic coefficients.

The magnetic field also requires that the duality transformation, Eq. (\ref{dualtrans}), be generalized to include
a component proportional to the original fields \cite{Dykhnemag,DR1994}. In the complex notation it reads
\be
\label{eq:comtran1}
{\cal J}'(\br)=\hat A {\cal J}(\br) -i\hat B {\cal F}(\br),~~~~~~
{\cal F}'(\br)=\hat C {\cal F}(\br) -i\hat D {\cal J}(\br),
\ee
where $\hat A,\hat B,\hat C, \hat D$ are $N\times N$ matrices that have to be real in order for the transformed
fields to still satisfy the conditions $\bnabla\cdot\bJ'_\alpha=0$ and $\bnabla\times\bF'_\alpha=0$.
Substituting Eqs. (\ref{eq:complexnot},\ref{eq:comtran1})
into ${\cal J}'(\br)=\hat{\cL}'(\br){\cal F}'(\br)$ yields the relation
\be
\label{eq:rel1}
\hat A\hat\cL(\br) - i\hat B=\hat\cL'(\br)\left[\hat C-i\hat D\hat\cL(\br)\right].
\ee
A similar relation
\be
\label{eq:rel1eff}
\hat A\hat\cL_{\rm eff} - i\hat B=\hat\cL'_{\rm eff}\left(\hat C-i\hat D\hat\cL_{\rm eff}\right).
\ee
holds for the effective response matrix $\hat\cL_{\rm eff}$ connecting the spatially averaged fields
$\langle{\cal J}\rangle=\hat\cL_{\rm eff}\langle{\cal F}\rangle$.

As before, we are interested in the transformation that fulfills Eq. (\ref{eq:rel1}) for the cases $\cL=\cL_1$, $\cL'=\cL_2$
and $\cL=\cL_2$, $\cL'=\cL_1$. Because $\cL_{1,2}$ are symmetric matrices one can verify that a solution to the first case
is also a solution to the second, provided that we take $\hat B$ and $\hat D$ to be symmetric and set $\hat C=-\hat A^T$.
Employing this choice and separating the real and imaginary parts of Eq. (\ref{eq:rel1}) we arrive at the defining
relations for the required duality transformation
\begin{eqnarray}
\label{eq:isorelr}
&&\!\!\!\!\!\!\!\!\!\hat A \hat L_1 + \hat L_2 \hat A^T + \hat L_2 \hat D \hat L_{H1} + \hat L_{H2} \hat D \hat L_1 = 0, \\
\label{eq:isoreli}
&&\!\!\!\!\!\!\!\!\!\hat B + \hat A \hat L_{H1} + \hat L_{H2} \hat A^T - \hat L_2 \hat D \hat L_1 + \hat L_{H2} \hat D \hat L_{H1} = 0.
\end{eqnarray}
We note that these equations constitute only $2N^2$ conditions for the $2N^2+N$ independent entries in
$\hat A$, $\hat B$ and $\hat D$, thereby leaving $N$ of them undetermined. Nevertheless, we find that
the resulting freedom in choosing the duality transformation does not manifest itself in the effective response
matrix. The latter is obtained, once $\hat A$, $\hat B$ and $\hat D$ have been established, by solving
Eq. (\ref{eq:rel1eff}) with $\hat\cL'_{\rm eff}=\hat\cL_{\rm eff}=\hat L_{\rm eff}-i\hat L_{H{\rm eff}}$, which
translates to solving
\begin{eqnarray}
\label{eq:isorelreff}
&&\!\!\!\!\!\!\!\!\!\hat A \hat L_{\rm eff} + \hat L_{\rm eff} \hat A^T + \hat L_{\rm eff} \hat D \hat L_{H{\rm eff}}
+ \hat L_{H{\rm eff}} \hat D \hat L_{\rm eff} = 0, \\
\label{eq:isorelieff}
&&\!\!\!\!\!\!\!\!\!\hat B + \hat A \hat L_{H{\rm eff}} + \hat L_{H{\rm eff}} \hat A^T - \hat L_{\rm eff} \hat D \hat L_{\rm eff}
+ \hat L_{H{\rm eff}} \hat D \hat L_{H{\rm eff}} = 0.
\end{eqnarray}

\subsection{The thermoelectric response of a two-phase single-layer in a magnetic field}

The above scheme can be used to numerically calculate $\hat \cL_{\rm eff}$ for a general two-phase layer.
We were able to obtain closed analytical results for the thermoelectric response in the simple case where
the two phases differ only by the sign of the charge carriers, implying
\be
\label{phmodel}
\hat L_{1,2}= \left( \begin{array}{cc} \sigma & \pm\alpha \\ \pm\alpha & \kappa \end{array} \right),~~~
\hat L_{H1,H2}= \left( \begin{array}{cc} \pm\sigma_H & \alpha_H \\ \alpha_H & \pm\kappa_H \end{array} \right).
\ee
In this case $\hat B$ and $\hat D$ are diagonal matrices, and the latter can be chosen as
\be
\hat D=\left(\begin{array}{cc} a & 0 \\
0 & b \end{array} \right),
\ee
with $a$ and $b$ arbitrary constants of dimensions $[\sigma^{-1}]$ and $[\kappa^{-1}]$, respectively, reflecting
the aforementioned freedom. In terms of them and the Hall angles
\be
\label{Hallangles}
\tan\theta_\sigma=\frac{\sigma_H}{\sigma},~~~\tan\theta_\kappa=\frac{\kappa_H}{\kappa},~~~
\tan\theta_\alpha=\frac{\alpha_H}{\alpha},
\ee
defined in the range $[-\pi/2,\pi/2]$, the matrix $\hat A$ takes the form
\be
\hat A=\frac{\alpha}{\sin(\theta_\sigma+\theta_\kappa)}
\left( \begin{array}{cc}0 & \hspace{0cm}\frac{\sigma}{\kappa}\frac{\cos\theta_\kappa}{\cos\theta_\sigma}a
-\frac{\cos(\theta_\sigma+\theta_\kappa-\theta_\alpha)}{\cos\theta_\alpha} b \\
\frac{\kappa}{\sigma}\frac{\cos\theta_\sigma}{\cos\theta_\kappa}b
-\frac{\cos(\theta_\sigma+\theta_\kappa-\theta_\alpha)}{\cos\theta_\alpha} a & \hspace{0cm} 0 \end{array} \right),
\ee
while $\hat B$ is given by Eq. (\ref{eq:isoreli}). Substituting the transformation back into Eqs. (\ref{eq:isorelreff},\ref{eq:isorelieff})
and solving for $\hat \cL_{\rm eff}$ yields 16 solutions. All are independent of $a$ and $b$ but only one produces real
transport coefficients that are also consistent with the physical requirements $\sigma_{\rm eff}\geq 0$ and $\kappa_{\rm eff}\geq 0$.
This solution is
\begin{eqnarray}
&&\sigma_{\rm eff}=\frac{\sigma}{\cos\theta_\sigma}\sqrt{1-\frac{\alpha^2}{\sigma\kappa}
\frac{\cos\theta_\sigma\cos\theta_\kappa}{\cos^2\left(\frac{\theta_\sigma+\theta_\kappa}{2}\right)}},\\
&&\kappa_{\rm eff}=\frac{\kappa}{\cos\theta_\kappa}\sqrt{1-\frac{\alpha^2}{\sigma\kappa}
\frac{\cos\theta_\sigma\cos\theta_\kappa}{\cos^2\left(\frac{\theta_\sigma+\theta_\kappa}{2}\right)}},\\
&&\alpha_{H{\rm eff}}=\alpha\left[\tan\theta_\alpha-\tan\left(\frac{\theta_\sigma+\theta_\kappa}{2}\right)\right], \\
&&\alpha_{\rm eff}=\sigma_{H{\rm eff}}=\kappa_{H{\rm eff}}=0,
\end{eqnarray}
which agrees with Eq. (\ref{LefflayerB0}) when applied to model (\ref{phmodel}) in the limit of vanishing Hall angles.

\subsection{The thermoelectric response of a two-phase double-layer in a magnetic field}

Obtaining a complete analytical solution to Eqs. (\ref{eq:isorelr})-(\ref{eq:isorelieff}) becomes difficult
even for the simplest models when the number of conserved currents is increased beyond two. However, one can
treat the problem perturbatively in the magnetic field, $H$, as we next demonstrate for the double-layer model
defined in Eq. (\ref{Lphmodelbi}) and augmented by the Hall part of the thermoelectric response of the two phases
\be
\label{LphmodelbiH}
\hat L_{H1,H2}= \left(\begin{array}{ccc}
\pm\sigma_H & \pm\eta_H & \alpha_H  \\
\pm\eta_H & \pm\sigma_H & \alpha_H \\
\alpha_H & \alpha_H & \pm 2\kappa_H  \end{array} \right).
\ee

To proceed we assume that $\hat L_{1,2}$ are
independent of $H$ and $\hat L_{H1,H2}$ are linear in $H$. Based on our knowledge of the field-free case
we look for a solution where $\hat B$, $\hat D$ and $\hat L_{\rm eff}$ are even functions of $H$ while
$\hat A$ and $\hat L_{H{\rm eff}}$ are odd, implying the expansions $\hat A= \hat A^{(1)} + \hat A^{(3)} + \dots$,
and $\hat B= \hat B^{(0)} + \hat B^{(2)} + \dots$, etc. Plugging these expansions into Eqs. (\ref{eq:isorelr},\ref{eq:isoreli})
one finds at zeroth order
\be
\label{eq:t01}
\hat B^{(0)}=\hat L_2\hat D^{(0)}\hat L_1.
\ee
This equation leaves 3 parameters undetermined, which we denote by $a,b,z$ and incorporate into the form of $\hat D^{(0)}$
\be
\label{D0}
\hat D^{(0)}= \left(\begin{array}{ccc}
a+b & a-b & z  \\
a-b & a+b & z \\
z & z & \frac{\sigma+\eta}{\kappa}a  \end{array} \right).
\ee
$\hat B^{(0)}$ is then given by Eq. (\ref{eq:t01}), and the zeroth order contribution to $\hat L_{\rm eff}$ is
determined from the corresponding order of Eq. (\ref{eq:isorelieff})
\be
\hat B^{(0)}=\hat L^{(0)}_{\rm eff}\hat D^{(0)}\hat L^{(0)}_{\rm eff}.
\ee
This equation yields 8 solutions, of which one is physical and properly
agrees with Eqs. (\ref{Leff0i})-(\ref{Leff0f}).

The first order equation, which determines $A^{(1)}$ and derived from Eq. (\ref{eq:isorelr})
\be
\label{eq:t11}
\hat A^{(1)}\hat L_1+\hat L_2\left.{\hat A}^{(1)}\right.^{T}\!\! +\hat L_2 \hat D^{(0)} \hat L_{H1}
+\hat L_{H2} \hat D^{(0)} \hat L_1=0,
\ee
has a solution provided we set $z=0$ in $\hat D^{(0)}$ (and consequently in $\hat B^{(0)}$).
The solution includes a new undetermined constant, $c$, and can be written in the form
\be
\label{A1}
\hat A^{(1)}= \left(\begin{array}{ccc}
0 & 0 & \frac{\sigma+\eta}{\kappa}\left[\left(\frac{\sigma_H+\eta_H}{\sigma+\eta}+\frac{\kappa_H}{\kappa}
-2\frac{\alpha_H}{\alpha}\right)\alpha a -\frac{c}{2}\right]  \\[5pt]
0 & 0 & \frac{\sigma+\eta}{\kappa}\left[\left(\frac{\sigma_H+\eta_H}{\sigma+\eta}+\frac{\kappa_H}{\kappa}
-2\frac{\alpha_H}{\alpha}\right)\alpha a -\frac{c}{2}\right] \\[5pt]
c & c & 0 \end{array}\right).
\ee
Substituting this result into Eq. (\ref{eq:isorelreff}) and expanding to first order yields an equation
that is the same as Eq. (\ref{eq:t11}) upon substituting $\hat L_1=\hat L_2=\hat L^{(0)}_{\rm eff}$ and
$\hat L_{H1}=\hat L_{H2}=\hat L^{(1)}_{H{\rm eff}}$. Solving for the latter gives the first order contribution
to the effective Hall components
\begin{eqnarray}
\label{LeffH1}
&&\sigma^{(1)}_{H{\rm eff}}=0,~~~~\eta^{(1)}_{H{\rm eff}}=0,~~~~\kappa^{(1)}_{H{\rm eff}}=0, \\
&&\alpha^{(1)}_{H{\rm eff}}=\alpha_H-\frac{\alpha}{2}\left(\tan\theta_{\sigma_+}+\tan\theta_\kappa \right),
\end{eqnarray}
where for brevity we have introduced
\be
\sigma_{\pm}=\sigma\pm\eta, ~~~~~~ \tan\theta_{\sigma_\pm}=\frac{\sigma_H\pm\eta_H}{\sigma\pm\eta}.
\ee

Finally, the second order equation derived from Eq. (\ref{eq:isoreli})
\be
\label{eq:t21}
\hat B^{(2)} + \hat A^{(1)} \hat L_{H1} + \hat L_{H2}\!\left.{\hat A}^{(1)}\right.^{\!T}
- \hat L_2 \hat D^{(2)} \hat L_1 + \hat L_{H2} \hat D^{(0)} \hat L_{H1} = 0,
\ee
may be solved for $\hat B^{(2)}$ and $\hat D^{(2)}$. The solution, which depends on 3 additional free parameters
is then substituted into Eq. (\ref{eq:isorelieff}) and results in
\begin{eqnarray}
\nonumber
&&\!\!\!\hspace{-0.5cm} \hat B^{(2)} + \hat A^{(1)} L_{H{\rm eff}}^{(1)} + L_{H{\rm eff}}^{(1)}\left.{\hat A}^{(1)}\right.^{\!T}
= \hat L_{\rm eff}^{(2)} \hat D^{(0)} \hat L_{\rm eff}^{(0)} \\
&&\!\!\!\hspace{-0.5cm} + \hat L_{\rm eff}^{(0)} \hat D^{(2)} \hat L_{\rm eff}^{(0)}
 + \hat L_{\rm eff}^{(0)} \hat D^{(0)} \hat L_{\rm eff}^{(2)}
-  L_{H{\rm eff}}^{(1)} \hat D^{(0)} L_{H{\rm eff}}^{(1)}.
\end{eqnarray}
This in turn yields the second order corrections contained in $\hat L_{\rm eff}^{(2)}$
\begin{eqnarray}
&&\hspace{-1.2cm}\sigma_{\rm eff}^{(2)}=\frac{\sigma_+}{4}\cos\theta
\left[\tan^2\theta_{\sigma_+} + \frac{\tan^2\theta}{4}\left(\tan\theta_{\sigma_+}-\tan\theta_\kappa\right)^2
\right]+\frac{\sigma_-}{4}\tan^2\theta_{\sigma_-}, \\
&&\hspace{-1.2cm}\eta_{\rm eff}^{(2)}=\sigma_{\rm eff}^{(2)}-\frac{\sigma_-}{2}\tan^2\theta_{\sigma_-}, \\
&&\hspace{-1.2cm}\alpha_{\rm eff}^{(2)}=0, \\
&&\hspace{-1.2cm}\kappa_{\rm eff}^{(2)}=\kappa\cos\theta
\left[\tan^2\theta_{\kappa} + \frac{\tan^2\theta}{4}\left(\tan\theta_{\sigma_+}-\tan\theta_\kappa\right)^2\right],
\end{eqnarray}
which are again invariant with respect to the freedom in the duality transformation.

\section{Discussion}

In this work we have provided new exact solutions to the two-phase model in multi-layer systems with both electrical
and heat currents. Our analysis relies on several assumptions. First, only the linear response regime is considered.
Second, that the systems is electrically and thermally isolated. Third, we assume strong thermal coupling between layers
in a multi-layered system but no electrical current leakage.  Obviously, these assumptions are
an idealization of reality, and cease to hold true beyond the relaxation lengths set by inter-layer electrical leakage
and thermal coupling to the environment. Nevertheless, our results are relevant on scales shorter than these relaxation
lengths, provided that they are much larger than the typical inhomogeneity scale.

The ability to obtain an exact solution to the problem crucially depends on the statistical equivalence between the
spatial distributions of the two phases. Violating this condition, \eg, by moving away from the percolation critical
point, necessitates a perturbative approach or a numerical solution. Our results provide benchmarks for the latter.

\section*{Acknowledgements}
This research was supported by the Israel Science Foundation (ISF) Grant No. 302/14 (O.A.) and Grant No. 701/17 (D.O.),
by the Binational Science Foundation (BSF) Grant No. 2014265 (D.O.), and by the Simons Foundation (I.A.).

\appendix

\section{}

\section*{Derivation of Eqs. (\ref{LefflayerB0},\ref{cdef})}

Making use of the fact that $\hat L_{1,2}$, and therefore $\hat U=\hat L_{\rm eff}$, are symmetric matrices
we expand them as
\be
\hat U= u+ \hat{u},~~~~~~\hat L_j= l_j+ \hat{l}_j,
\ee
where $u= u_0 I$ and $\hat{u}= u_x \hat\sigma_x+ u_z \hat\sigma_z$. Here, $I$ is the unit matrix, $\hat\sigma_{x,z}$
are the Pauli matrices and $u_\mu$ are constants. Similar definitions hold for $l$ and $\hat l$.
Equation (\ref{rotation}) then implies
\be
\label{2b2expand}
\left\{l_2+ \hat{l}_2, \frac{1}{\sqrt{d_1 d_2}}(u -\hat{u})\right\}=\left\{u+ \hat{u}, \frac{1}{d_1}( l_1-\hat{l}_1)\right\},
\ee
where $\{\cdot,\cdot\}$ is the anticommutator and where we have defined $d_j=\det \hat L_j$ and used the relation
\be
\det \hat U= \sqrt{d_1 d_2}, \label{det2}
\ee
which follows from Eq.~(\ref{rotation}). In turn, Eq. (\ref{2b2expand}) yields
\begin{eqnarray}
\nonumber
&& u\left(\frac{\hat l_1}{\sqrt{d_1}}+\frac{\hat l_2}{\sqrt{d_2}}\right)=\left(\frac{l_1}{\sqrt{d_1}}+\frac{l_2}{\sqrt{d_2}}\right)\hat u , \\
&& 2u\left(\frac{l_1}{\sqrt{d_1}}-\frac{l_2}{\sqrt{d_2}}\right) = \left\{ \frac{\hat l_1}{\sqrt{d_1}}-\frac{\hat l_2}{\sqrt{d_2}},\hat u\right\},
\end{eqnarray}
whose solution is readily obtained as Eqs. (\ref{LefflayerB0},\ref{cdef}).

\section*{The solution of the double-layer problem without magnetic field}

Here we detail the solution of Eq. (\ref{rotation}) for the double-layer model defined by Eq. (\ref{eq:doublelayer}).
We begin by representing the matrix $\hat{U}= \hat{L}_{eff}$ in the form
\begin{equation}
\hat{U}= \left( \begin{array}{ccc} u_1 & u_2 &u_4 \\ u_2 & u_1 & u_4 \\ u_4 & u_4 &  u_3 \end{array} \right).
 \end{equation}
Next, noticing that $\det^2(\hat{U}) =\det \hat{L}_1 \det \hat{L}_2$ we rewrite Eq. (\ref{rotation}) as
\begin{equation}
\label{eq:Q}
\hat{Q}\equiv\beta \det( \hat{L}_1)\, U \hat{L}_1^{-1} - \det(\hat{U})\, \hat{L}_2 \hat{U}^{-1}=0,
\end{equation}
where $\beta=(\det \hat{L}_2/\det \hat{L}_1)^{1/2}$.

The solution of  Eq. (\ref{eq:Q}) is now obtained by the following steps. First we solve $Q_{33}=0$ for $u_3$,
\begin{equation}
u_3=\frac{2}{\beta}\frac{ (u_1-u_2)u_4 \alpha_2 +(u_2^2-u_1^2)\kappa_2 +u_1 \alpha_1 (\eta_1-\sigma_1)\beta}{ (\eta_1^2
-\sigma_1^2)}.
\end{equation}
Next we substitute $u_3$ in the equation $Q_{13}=0$ and solve for $u_4$,
\begin{equation}
u_4=\frac{(u_1+u_2)[(u_1-u_2) \alpha_2  +\alpha_1(\sigma_1-\eta_1)\beta]}{\beta(\sigma_1^2- \eta_1^2)+ (u_1-u_2) (\eta_2 +\sigma_2)}.
\end{equation}
We substitute the above expressions for $u_3$ and $u_4$ in the equation $Q_{11}+ Q_{12}=0$ and solved it for $u_2$,
\begin{equation}
u_2= u_1-\sqrt{ (\sigma_1-\eta_1)(\sigma_2 -\eta_2)}.
\end{equation}
Finally, substituting the above expressions in the equation $Q_{11}=0$ turns it into a quadratic equation for $u_1$,
whose physical root ($u_1>0$) completes the general solution for the model.

\end{document}